\documentclass[runningheads]{llncs}
\usepackage[T1]{fontenc}
\usepackage{url}
\usepackage{adjustbox}

\usepackage{graphicx}
\usepackage{amsmath,amsfonts,amssymb}
\begin{document}
\title{Performance Estimation for Supervised Medical Image Segmentation Models on Unlabeled Data Using UniverSeg}
\titlerunning{Estimating Unseen Data Performance in Medical Image Segmentation}
\author{Jingchen Zou \inst{1} \and
Jianqiang Li \inst{1} \and
Gabriel Jimenez \inst{2} \and
Qing Zhao \inst{1} \and
Daniel Racoceanu \inst{3} \and
Matias Cosarinsky \inst{4} \and
Enzo Ferrante \inst{4} \and
Guanghui Fu \inst{3}
}
\authorrunning{J. Zou et al.}
\institute{College of Computer Science, Beijing University of Technology, Beijing, China \and % 1
Sorbonne Université, Institut du Cerveau - Paris Brain Institute - ICM, CNRS, Inserm, AP-HP, Hôpital de la Pitié Salpêtrière, Paris, France \and % 2
Sorbonne Université, Institut du Cerveau - Paris Brain Institute - ICM, CNRS, Inria, Inserm, AP-HP, Hôpital de la Pitié Salpêtrière, Paris, France \and % 3
Instituto de Ciencias de la Computación, CONICET - Universidad de Buenos Aires, Argentina \\ % 4
\email{guanghui.fu@icm-institute.org}}
\maketitle              % typeset the header of the contribution
\begin{abstract}
The performance of medical image segmentation models is usually evaluated using metrics like the Dice score and Hausdorff distance, which compare predicted masks to ground truth annotations. However, when applying the model to unseen data, such as in clinical settings, it is often impractical to annotate all the data, making the model's performance uncertain. To address this challenge, we propose the Segmentation Performance Evaluator (SPE), a framework for estimating segmentation models' performance on unlabeled data. 
% In the SPE framework, segmentation models saved at different training epochs to generate predicted masks on the test set, capturing their varying performance. The generated predicted masks can serve as the support set for UniverSeg. UniverSeg then performs inference on the training set, and pseudo-performance is calculated by comparing the predictions with ground-truth labels. A linear function maps pseudo-performance to real performance. In practice, this function estimates the model's performance from the pseudo-performance.
This framework is adaptable to various evaluation metrics and model architectures. Experiments on six publicly available datasets across six evaluation metrics including pixel-based metrics such as Dice score and distance-based metrics like HD95, demonstrated the versatility and effectiveness of our approach, achieving a high correlation (0.956$\pm$0.046) and low MAE (0.025$\pm$0.019) compare with real Dice score on the independent test set. These results highlight its ability to reliably estimate model performance without requiring annotations.
The SPE framework integrates seamlessly into any model training process without adding training overhead, enabling performance estimation and facilitating the real-world application of medical image segmentation algorithms.
The source code is publicly available at: \url{https://anonymous.4open.science/r/SPE}.
\end{abstract}
\section{Introduction} \label{sec:intro} 

Automatic medical image segmentation is a critical task in image analysis frameworks ~\cite{azad2024medical,chen2022recent}. Segmenting lesions or anatomical structures supports clinical decision-making, such as surgical planning~\cite{scorza2021surgical} or disease characterization~\cite{wasserthal2023totalsegmentator}. Currently, supervised deep learning method serves as the foundation for constructing segmentation models, which require training on annotated images of target regions~\cite{wang2021annotation,mo2022review}. Representative architectures like UNet~\cite{ronneberger2015u} can achieve high accuracy when provided with sufficient image-label pairs during training~\cite{azad2024medical}.
However, estimating model performance on unseen clinical data remains challenging~\cite{dinsdale2022challenges}. While visual inspection or annotating additional image-label pairs is a common approach, it is impractical for large cohorts due to the high cost and time requirements of manual annotation. Even with extensive annotations, models will inevitably encounter unseen data. Therefore, an automatic estimation framework is essential for real-world applications.

Vanya et al.~\cite{rca} proposed a novel framework named as reverse classification accuracy (RCA). It trains a reverse classifier using the predicted segmentation from a new image and evaluates it on reference images with ground truth. A high-quality prediction leads to good reverse classifier performance on some reference images. 
However, RCA is primarily suited for atlas-based segmentation models or anatomical structures with minimal variation, making it ineffective for lesion segmentation due to altered anatomy.
The development of UniverSeg~\cite{butoi2023universeg}, a foundation model for medical imaging, offers a flexible tool that can be explored for performance estimation. UniverSeg segments new images by referencing a support set of image-label pairs and adapts to varying anatomical and pathological conditions. Leveraging the strong correlation between support set quality and segmentation performance, this approach provides a promising direction for estimating model performance, extending beyond the limitations of RCA.

In this paper, we propose a flexible performance estimation framework called the Segmentation Performance Evaluator (SPE).
In the SPE framework, segmentation models are saved at different training epochs to capture their varying performance levels on the test set, which are considered as the real performance. These models generate predicted masks on the test set at each epoch. UniverSeg then uses these predicted masks as the support set to define its tasks and performs inference on images from the training set. Pseudo-performance is calculated by comparing the newly generated predictions with the ground truth in the training set. A simple linear function is fitted to map the pseudo-performance to the corresponding real performance. This linear mapping is used during real-world applications to estimate the model's actual performance based on pseudo-performance.
We conducted experiments on six datasets across five imaging modalities, covering lesions and anatomical structures, to estimate six pixel- and distance-level metrics.
The results shown its effectiveness, with a high correlation and low MAE on the independent dataset, highlighting its ability to estimate model performance without annotations.
This framework integrates seamlessly into the training process without restrictions on model architecture, enabling performance estimation for clinical applications of medical image segmentation algorithms.
All source codes are publicly available.

\section{Methods} \label{sec:methods}
\begin{figure}[!hbtp]
\centering
\includegraphics[width=0.8\linewidth]{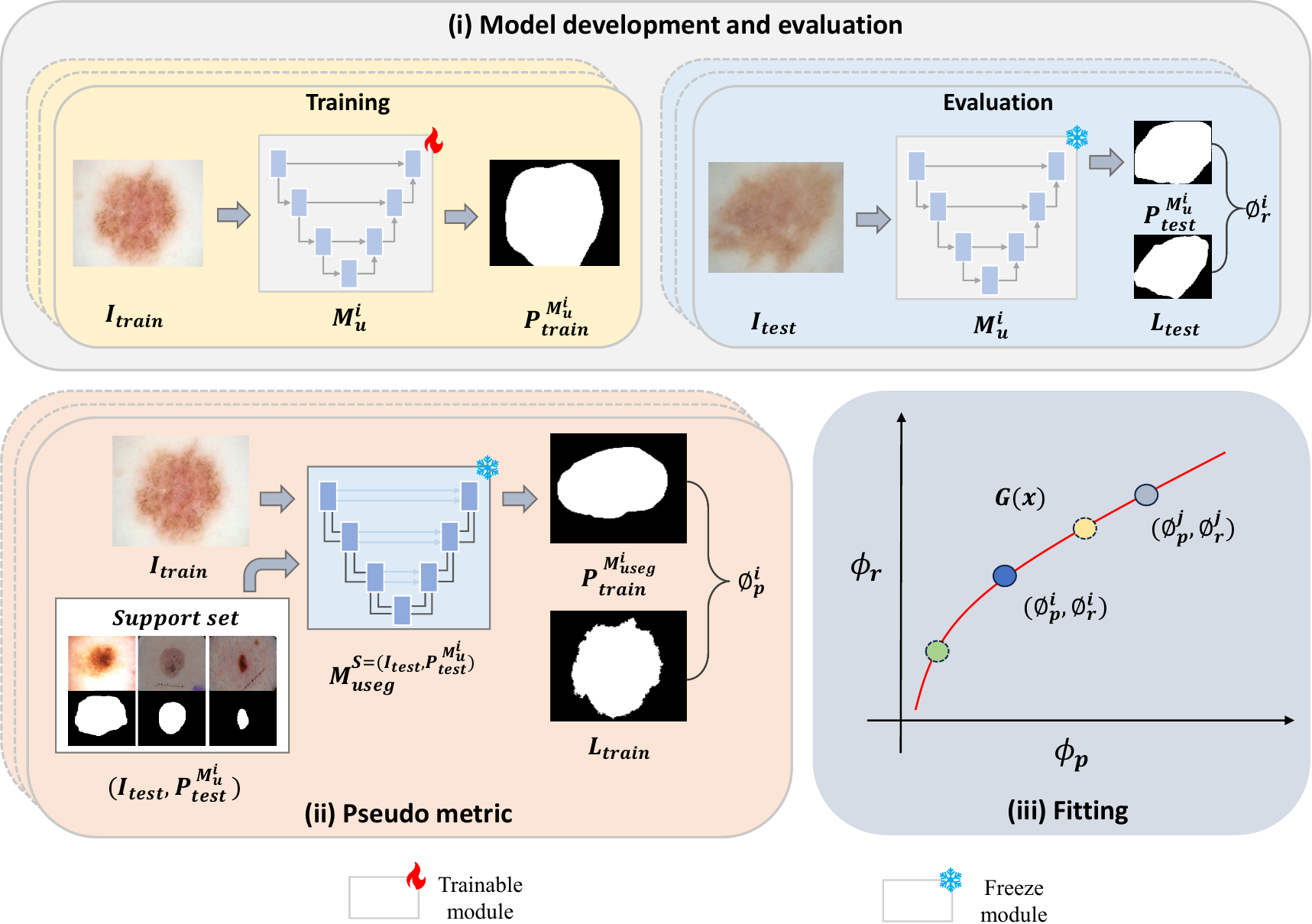}
\caption{An overview of our proposed SPE framework.}
\label{fig:overall_framework}
\end{figure}

The SPE framework consists of four stages: the training stage, inference stage, pseudo-metric computation, and fitting stage. The framework process is illustrated in Figure~\ref{fig:overall_framework}.  

\subsection{Model development and evaluation}
% \subsection{Model training}
\paragraph{Training} In a standard deep learning-based segmentation model development process, the annotated dataset ($I$ for images and $L$ for corresponding labels) is typically divided into three subsets: a training set ($I_{\text{train}}$ and $L_{\text{train}}$), a validation set, and a test set ($I_{\text{test}}$ and $L_{\text{test}}$). The training and validation sets are used for model development, while the test set is reserved for evaluation.
The segmentation model is not restricted within our framework; we use UNet ($M_u$) as an example since it is a standard model in the field. We assume the model is trained for $K$ epochs, with the model trained at the $i$-th epoch denoted as $M_u^i$.

% \subsection{Model testing stage}
\paragraph{Evaluation} During the evaluation stage, the model $M_u^i$ is used to infer segmentation results $P_{\text{test}}^{M_u^i}$ from the test set images $I_{\text{test}}$.
Given the annotated labels $L_{\text{test}}$ in the test set, the real performance of $M_u^i$ can be calculated as:
\begin{equation}
    \phi_r^i = F(L_{\text{test}}, P_{\text{test}}^{(M_u^i)})
\end{equation}
where $F$ denotes any segmentation evaluation metric function, such as the Dice score or Hausdorff distance.

% \subsection{Estimator backbone: UniverSeg}
% Long version
% UniverSeg\cite{butoi2023universeg} efficiently tackles unseen medical segmentation tasks without additional training, making it suitable for clinical use. Trained on 53 publicly available datasets with over 22,000 scans covering diverse anatomical structures and imaging modalities, it uses a support set of image-label pairs to infer segmentation on query images. Through its CrossBlock mechanism, UniverSeg leverages relationships between the query and support set for effective cross-task generalization. Experimental results show that the quality of the support set directly impacts segmentation performance on new data.
\subsection{Reverse pseudo metric calculation}
UniverSeg\cite{butoi2023universeg} is a Foundation Model for medical image segmentation that addresses unseen tasks without additional training, making it suitable for clinical use. It relies on a support set of image-label pairs to infer query image segmentation. Segmentation performance is directly tied to the quality of the support set: good support sets enhance performance, while poor ones lead to degradation.

In the reverse metric computation stage, we use the UniverSeg model $M_{\text{useg}}^S$, configured with a support set $S = (I_{\text{test}}, P_{\text{test}}^{M_u^i})$. 
The reason we use the test set rather than the validation set is that the validation set is typically used for tuning model performance, whereas the reverse pseudo-metric calculation process requires an independent dataset to ensure unbiased mapping.
This model is then used to perform inference on the training set $I_{\text{train}}$, which has annotated labels available.
The UniverSeg model $M_{\text{useg}}^S$ generates segmentation results $P_{\text{train}}^{M_{\text{useg}}}$. These results are compared with the true labels $L_{\text{train}}$ of the training set, and the pseudo performance metric $\phi_p^i$ is calculated as follows:
\begin{equation}
    \phi_p^i = F(L_{\text{train}}, P_{\text{train}}^{M_{\text{useg}}^i})
\end{equation}

Since the model $M_u$ is trained for $K$ epochs during the training stage, these above stages can produce $K$ pairs of $(\phi_r^i, \phi_p^i)$ at epoch $i$. Here, $\phi_r^i$ represents the real performance of $M_u^i$ on the test set, and $\phi_p^i$ represents the pseudo performance metric from the reverse evaluation stage.
The pair of real performance and reverse pseudo performance is denoted as $\Psi$ and it represents performance of an entire group of images, rather than the individual image performance.
\begin{equation}
    \Psi = \{(\phi_r^i, \phi_p^i)\}_{i=1}^K
\end{equation}

\subsection{Performance linear function fitting}
In the fitting stage, our goal is to find an appropriate mapping function $G(x)$ that constructs the relationship between the estimate-performance metric and the real performance metric by minimizing the fitting error. The fitting process of the mapping function $G(x)$ can be formulated as an optimization problem:
\begin{equation}
    \min_G L(G) = \sum_{i=1}^K \left| \phi_r^i - G(\phi_p^i) \right|^2
    \label{eq:fit_func}
\end{equation}
where $L(G)$ is the loss function, representing the fitting error of the mapping function $G(x)$.

\subsection{Unlabeled data performance estimation}
After training the segmentation model, a model meeting the desired performance criteria (denoted as $M_u^d$) is selected for deployment in real-world scenarios. Since practical datasets ($I_{\text{ext}}$) typically lack real segmentation labels ($L_{\text{ext}}$), the actual segmentation performance $\phi_r$ cannot be computed using annotation-based metrics $F$. In such cases, SPE can estimate the performance of $M_u^d$ on the unlabeled dataset $I_{\text{ext}}$. 
Using $M_u^d$, we perform inference on $I_{\text{ext}}$ to generate predicted masks $P_{\text{ext}}^{M_u^d}$. The support set is constructed as the pair $(I_{\text{ext}}, P_{\text{ext}}^{M_u^d})$, which is then used by UniverSeg $M_{\text{useg}}^{S=(I_{\text{ext}}, P_{\text{ext}}^{M_u^d})}$. UniverSeg performs inference on the training set to compute the reverse pseudo-performance metric $\phi_p^d$.
Using the mapping function $G(x)$ obtained in Equation~\ref{eq:fit_func}, the real performance $\phi_r^d$ of the model can be estimated from $\phi_p^d$ as:  
\begin{equation}
    \hat{\phi_r^d} = G(\phi_p^d)
\end{equation}
where $\hat{\phi_r^d}$ represents the estimated performance metric.

\section{Experiments} \label{sec:experiments}

\subsection{Datasets}
\begin{figure}[!hbtp]
\centering
\includegraphics[width=0.8\linewidth]{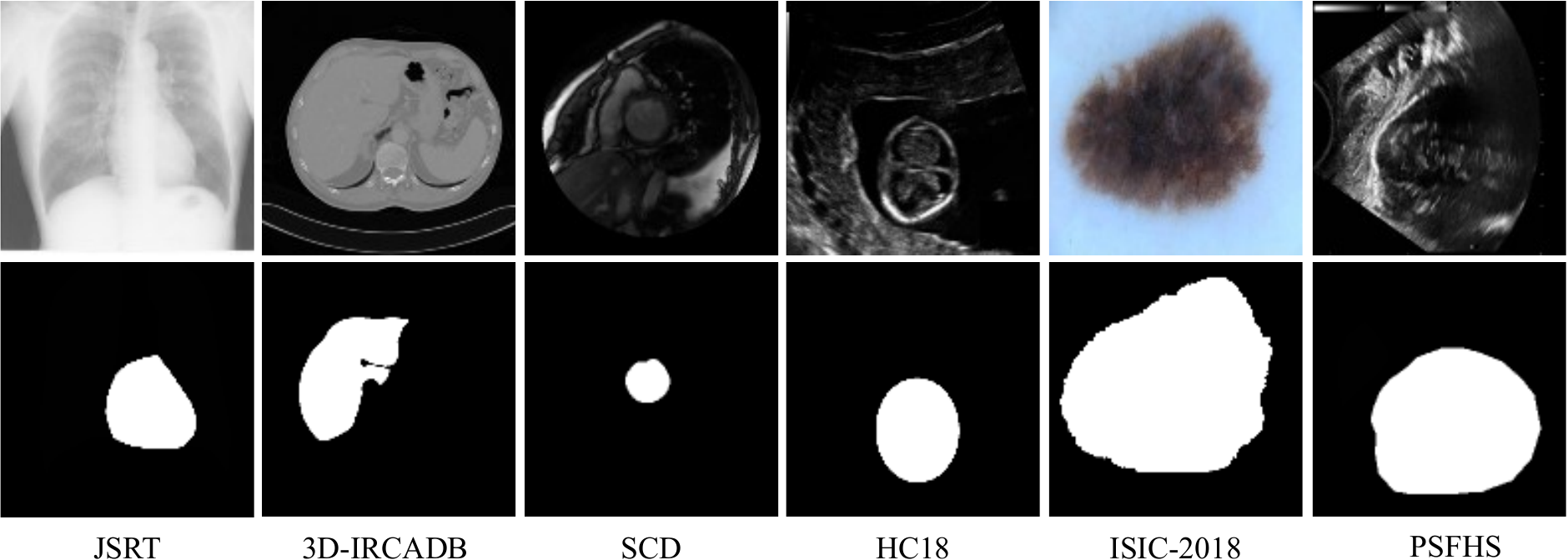}
\caption{Example images and their corresponding annotation masks from the experimental datasets.}
\label{fig:datasets}
\end{figure}

To thoroughly validate the generalization of the proposed SPE framework across different types of medical images, we conducted experiments on six medical image segmentation datasets. These datasets cover typical scenarios in various imaging modalities as shown in Figure~\ref{fig:datasets}. The data distribution can be seen in Table~\ref{tab:dataset}.
% % X-ray
The JSRT dataset~\cite{shiraishi2000development} consists of 246 chest X-ray images for lung region segmentation. 
% % CT
The 3D-IRCADB dataset~\cite{soler20103d} contains 20 abdominal contrast-enhanced CT scans in 3D volumetric format, enabling comprehensive evaluation of organ and tumor structures. 
% % MRIs
The SCD dataset~\cite{radau2009evaluation} includes 45 MRIs focused on left ventricle segmentation from the heart. 
% % Ultrasound-1
For fetal head segmentation, the HC18 dataset~\cite{van2018automated} provides 999 2D ultrasound images for measuring head circumference. 
% % Ultrasound-2
Additionally, the PSFHS dataset~\cite{chen2024psfhs} comprises 1,358 pixel-level annotated ultrasound images, for pubic symphysis and fetal head segmentation. 
% % Dermoscopic images
Lastly, the ISIC-2018 dataset~\cite{codella2019skin} includes 3,694 dermoscopic RGB images for melanoma segmentation.

During data preprocessing, 3D data is sliced into 2D images, retaining only those with masked regions of 20 pixels or more. All images are resized to 128$\times$128 to follow the requirement of UniverSeg.

\begin{table}[!ht]
    \centering
    \caption{Data distribution of the experimental datasets.}
    \begin{tabular}{cccccc}
    \hline
        \textbf{Dataset} & \textbf{Modality} & \textbf{Train} & \textbf{Validation} & \textbf{Test} & \textbf{Extra test} \\ \hline
        JSRT~\cite{shiraishi2000development} & X-ray & 137 & 35 & 49 & 25 \\
        3D-IRCADB~\cite{soler20103d} & CT & 1243 & 311 & 337 & 183 \\ 
        SCD~\cite{radau2009evaluation} & MRI & 450 & 113 & 161 & 81 \\
        HC18~\cite{van2018automated} & Ultrasound & 559 & 140 & 200 & 100 \\
        ISIC-2018~\cite{codella2019skin} & Dermoscopi & 2075 & 519 & 1000 & 100 \\ 
        PSFHS~\cite{chen2024psfhs} & Ultrasound & 760 & 190 & 272 & 136 \\ \hline
    \end{tabular}
    \label{tab:dataset}
\end{table}

\subsection{Evaluation metrics}
To evaluate the performance of SPE, we use the Mean Absolute Error (MAE), and the Pearson correlation coefficient, to measure the alignment between predicted and actual values.
MAE measures the deviation between the predicted performance metric $\hat{\phi}_r$ and the actual value $\phi_r$, reflecting the absolute error between SPE predictions and true values. A smaller MAE indicates a more accurate estimation as shown in Equation~\ref{eq:eval:mae}.
\begin{equation}
    \text{MAE} = \frac{1}{K} \sum_{i=1}^{K} \left| \phi_r^i - \hat{\phi_r^i} \right|
    \label{eq:eval:mae}
\end{equation}
Pearson correlation coefficient, evaluates the linear relationship between $\hat{\phi}_r$ and $\phi_r$. A correlation coefficient $\rho$ closer to 1 indicates stronger consistency between estimation and actual performance metrics shown in Equation~\ref{eq:eval:correla}.
\begin{equation}
    \text{Correlation} = \frac{\sum_{i=1}^{K} \left( \phi_r^i - \bar{\phi_r} \right) \left( \hat{\phi_r^i} - \overline{\hat{\phi_r}} \right)}{\sqrt{\sum_{i=1}^{K} \left( \phi_r^i - \bar{\phi_r} \right)^2 \sum_{i=1}^{K} \left( \hat{\phi_r^i} - \overline{\hat{\phi_r}} \right)^2}}
    \label{eq:eval:correla}
\end{equation}
where \(\bar{\phi_r}\) and \(\overline{\hat{\phi_r}}\) represent the mean of the true performance metrics \(\phi_r\) and the estimated performance metrics \(\hat{\phi_r}\), respectively.

\subsection{Implementation details}
Our code is built using the PyTorch framework~\cite{paszke2019pytorch}, with UNet~\cite{ronneberger2015u} as the segmentation model. The model is trained for 100 epochs using the Adam optimizer~\cite{kingma2014adam} with a learning rate of 1e-4. Model weights are saved every 5 epochs, resulting in 20 models with different weights.
During evaluation, we use the pre-trained UniverSeg model~\cite{butoi2023universeg} for inference, keeping its parameters frozen throughout. The support set size affects UniverSeg's inference speed and resource usage. For each $i$-th epoch, we randomly select 64 image pairs from $(I_{\text{test}}, P_{\text{test}}^{(M_u^i)})$, as UniverSeg supports a maximum of 64 images. Each experiment is repeated 6 times, and the average metric value is used as the final result.
We use SPE to estimate six widely used evaluation metrics: Dice score, HD95, Jaccard, Pearson correlation coefficient, Recall, and Precision.
All experimental code and trained models are publicly available for reproducibility\footnote{\url{https://anonymous.4open.science/r/SPE}}

\section{Results}
\begin{figure}[!hbtp]
\centering
\includegraphics[width=0.8\linewidth]{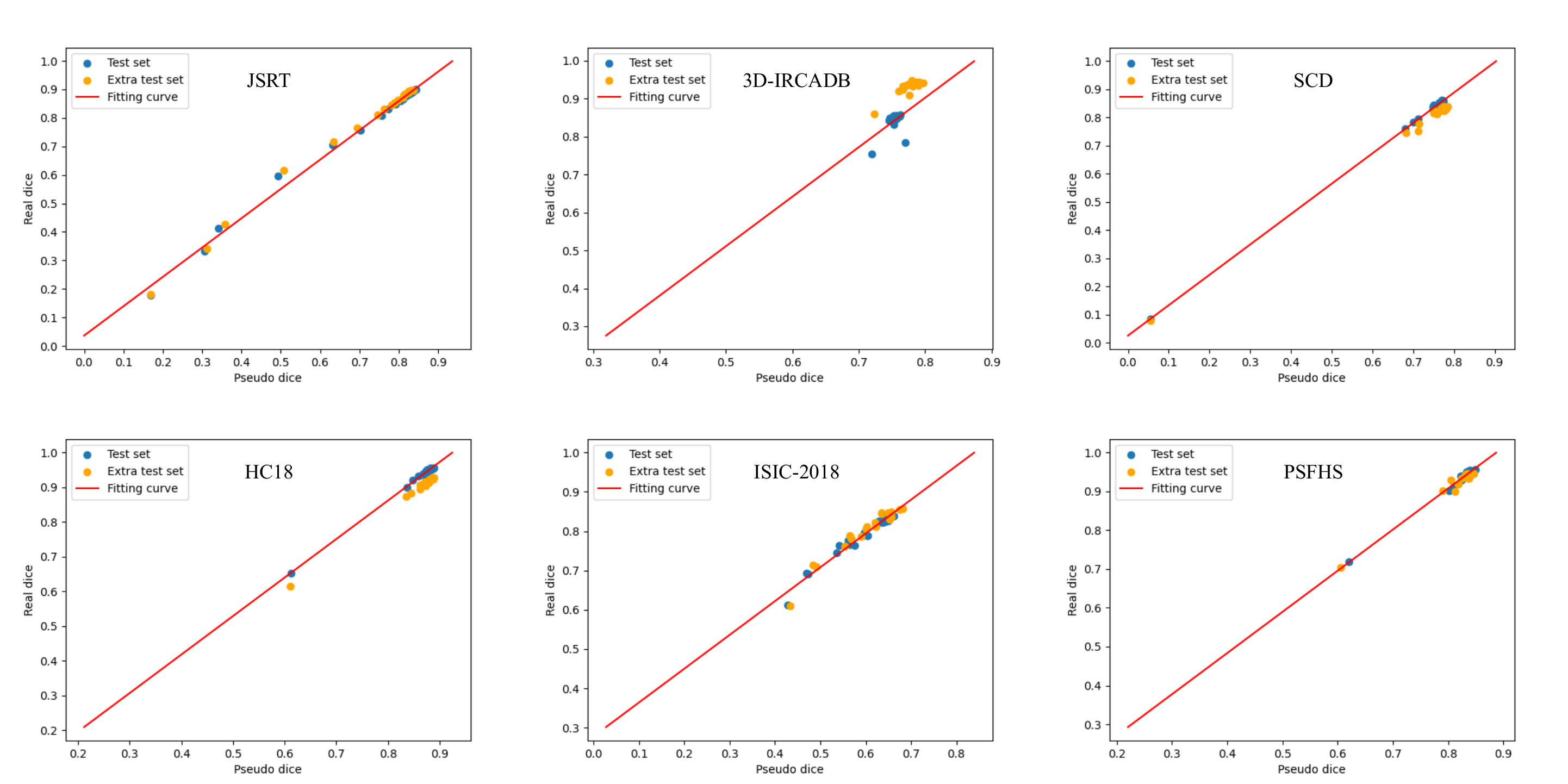}
\caption{SPE estimation of Dice metrics results on the six experimental datasets. }
\label{fig:result_dice}
\end{figure}

The estimation results are presented in Table~\ref{tab:result_total}, while Figure~\ref{fig:result_dice} illustrates the Dice score mapping function. The red curve shows the function $G(x)$, fitted using the test set, and the orange points correspond to the Extra test set, which evaluates generalization on previously unseen data. A tighter alignment of these points with the red curve indicates more accurate estimation. The result of other evaluation metrics can be seen in Appendix, Figure~\ref{fig:result_hd95},~\ref{fig:result_jaccard},~\ref{fig:result_pearsonsr},~\ref{fig:result_recall},~\ref{fig:result_precison}.

\begin{table}[ht]
\centering
\caption{The mean absolute error (MAE) and correlation (Corr) of SPE framework on various metrics across datasets.}
\begin{adjustbox}{max width=1\textwidth}
\begin{tabular}{c|cc|cc|cc|cc|cc|cc} 
\hline
\textbf{Dataset} & \multicolumn{2}{c|}{\textbf{Dice}} & \multicolumn{2}{c|}{\textbf{HD95}} & \multicolumn{2}{c|}{\textbf{Precision}} & \multicolumn{2}{c|}{\textbf{Recall}} & \multicolumn{2}{c|}{\textbf{Jaccard}} & \multicolumn{2}{c}{\textbf{Pearson}} \\ 
\cline{2-13}
                 & \textbf{MAE} & \textbf{Corr} & \textbf{MAE} & \textbf{Corr} & \textbf{MAE} & \textbf{Corr} & \textbf{MAE} & \textbf{Corr} & \textbf{MAE} & \textbf{Corr} & \textbf{MAE} & \textbf{Corr} \\ 
\hline
JSRT~\cite{shiraishi2000development} & 0.013 & 0.997 & 23.23 & 0.797 & 0.014 & 0.998 & 0.015 & 0.602 & 0.016 & 0.998 & 0.013 & 0.998 \\
3D-IRCADBS~\cite{soler20103d}        & 0.059 & 0.882 & 1.294 & 0.818 & 0.063 & 0.912 & 0.130 & 0.869 & 0.086 & 0.873 & 0.045 & 0.874 \\
SCD~\cite{radau2009evaluation}       & 0.025 & 0.999 & 1.349 & 0.999 & 0.042 & 0.998 & 0.029 & 0.778 & 0.044 & 0.999 & 0.041 & 0.999 \\
HC18~\cite{van2018automated}         & 0.031 & 0.998 & 1.386 & 0.994 & 0.034 & 0.997 & 0.016 & 0.498 & 0.033 & 0.998 & 0.010 & 0.999 \\
ISIC-2018~\cite{codella2019skin}     & 0.013 & 0.969 & 4.632 & 0.921 & 0.012 & 0.987 & 0.019 & 0.599 & 0.013 & 0.983 & 0.013 & 0.987 \\
PSFHS~\cite{chen2024psfhs}           & 0.007 & 0.991 & 1.294 & 0.988 & 0.008 & 0.991 & 0.010 & 0.857 & 0.009 & 0.991 & 0.008 & 0.992 \\
\hline
\textbf{Mean}                  & 0.025& 0.956 & 5.865& 0.919 & 0.029 & 0.981 & 0.037 & 0.701 & 0.033 & 0.957 & 0.022 & 0.958\\
\hline
\textbf{STD}                  & 0.019 & 0.046 & 8.687 & 0.081 & 0.021 & 0.036 & 0.044 & 0.160 & 0.028 & 0.054 & 0.016 & 0.045 \\
\hline

\end{tabular}
\end{adjustbox}
\label{tab:result_total}
\end{table}

The framework exhibited consistent performance with low MAE and high correlation values, underscoring its robustness and reliability.
For pixel-based metrics, such as the Dice score and precision, SPE achieved impressive performance. The Dice score estimation had an average MAE of 0.025$\pm$0.019 and a correlation of 0.956$\pm$0.046, reflecting its ability to closely align with true model performance. 
Similarly, precision exhibited an average MAE of 0.029$\pm$0.021 with a high correlation of 0.981$\pm$0.036, confirming the framework's reliability in estimating segmentation quality. The results for other metrics like recall and jaccard further validate the framework's versatility and adaptability across a wide array of segmentation evaluation metrics.
Distance-based metrics, like HD95, demonstrated the adaptability of SPE to complex evaluations. Despite the inherent variability in HD95 values across datasets, SPE maintained a mean MAE of 5.865$\pm$8.687 and a correlation of 0.919$\pm$0.081. This highlights the framework's capability to estimate performance even for metrics with wide ranges of values. The estimation on HD95 is lower than other pixel-level metrics due to its high sensitivity to outliers, as discussed further in Appendix, Sec~\ref{sec:appendix:linear}.

Dataset-specific results confirmed the generalizability of SPE. For example, the framework performed exceptionally well on the JSRT and PSFHS datasets, achieving MAE values as low as 0.007 and correlations nearing 0.999 for multiple metrics. On datasets like ISIC-2018 and 3D-IRCADBS, which encompass more challenging segmentation tasks, SPE maintained robust performance, demonstrating its utility across varying data characteristics.
These findings demonstrate SPE's ability to estimate segmentation performance without annotations, enabling its use in real-world clinical applications and settings with limited labeled data.

\section{Conclusion} \label{sec:conclusion}
In this paper, we introduced a novel framework for performance estimation on unlabeled data, leveraging the flexibility of the UniverSeg foundation model. Through experiments conducted on six evaluation metrics across six medical datasets spanning five imaging modalities, the proposed framework demonstrated accurate performance estimation, achieving high correlation and low MAE. Importantly, this approach integrates seamlessly into the training process, imposing no restrictions on model architecture and adding no computational overhead. These attributes make it a practical and adaptable solution for real-world applications in medical image segmentation.

% The proposed SPE framework offers a transformative solution to the challenge of evaluating segmentation models on unseen and unlabeled data, with significant implications for clinical and research applications. By leveraging the UniverSeg framework, SPE demonstrates robust adaptability across various imaging modalities, enabling accurate performance estimation without relying on annotated datasets. This is particularly valuable in clinical settings where labeled data is often scarce and difficult to obtain. Despite these strengths, certain limitations must be addressed to fully realize SPE's potential. Its performance variation on the HD95 metric (see Appendix A for details) suggests that more in-depth research is needed on different types of performance metrics, and that relying on simple linear fits may not always capture the complexity of real-world scenarios. Moreover, the computational demands of SPE highlight the need for optimization to ensure its scalability and accessibility in diverse environments. Future research should focus on refining the estimation methods, expanding the framework to broader imaging tasks, and evaluating its practical utility in real-world settings. By addressing these challenges and enhancing its capabilities, SPE has the potential to significantly advance the field of medical imaging, supporting the deployment of robust machine learning models and ultimately improving patient outcomes.

\section{Acknowledgments}\label{sec:acknowledgments}
This work was granted access to the HPC resources of IDRIS under the allocation 2023-AD011014513 made by GENCI.

\bibliography{refs} 
\bibliographystyle{splncs04}

\newpage

\appendix

\renewcommand\thefigure{S\arabic{figure}} 
\setcounter{figure}{0} \renewcommand\thetable{S\arabic{table}} 
\setcounter{table}{0}

\section{Performance estimation of other evaluation metrics} \label{sec:appendix:other_metrics}
\begin{figure}[!hbtp]
\centering
\includegraphics[width=1\linewidth]{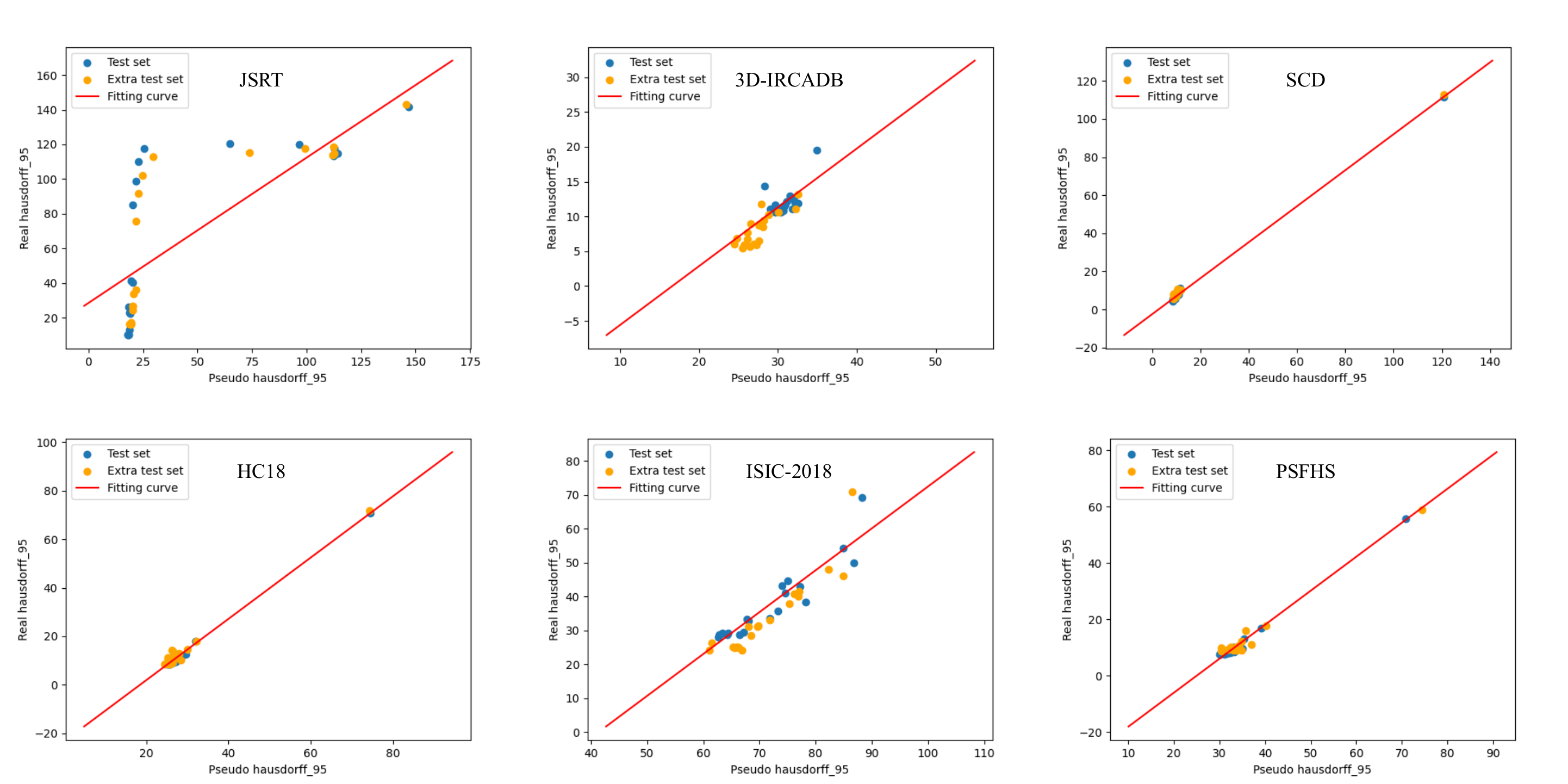}
\caption{SPE estimation of 95\% Hausdorff distance (HD95) metrics results. }
\label{fig:result_hd95}
\end{figure}

\begin{figure}[!hbtp]
\centering
\includegraphics[width=1\linewidth]{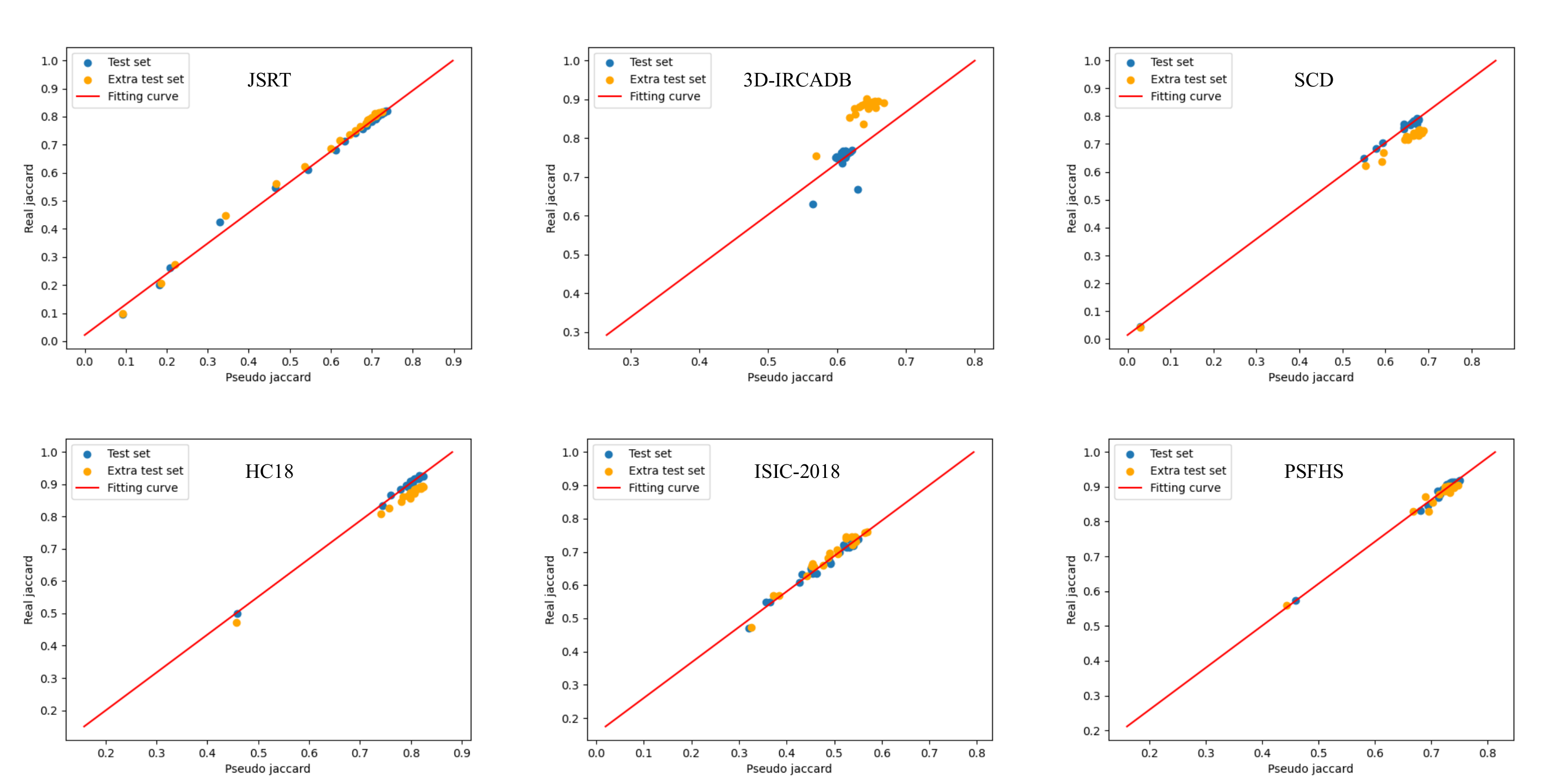}
\caption{SPE estimation of jaccard metrics results. }
\label{fig:result_jaccard}
\end{figure}

\begin{figure}[!hbtp]
\centering
\includegraphics[width=1\linewidth]{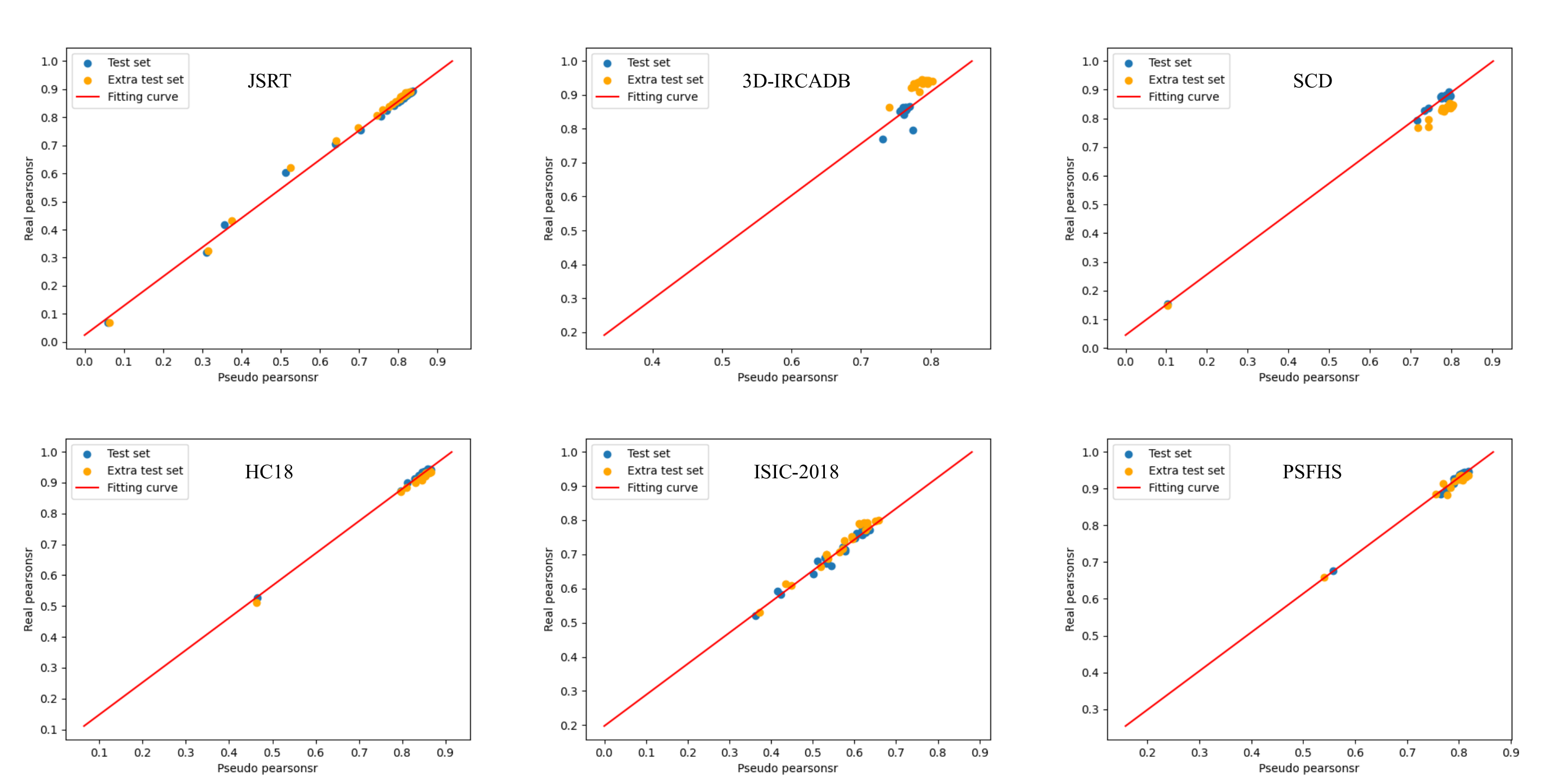}
\caption{SPE estimation of Pearson correlation coefficient metrics results. }
\label{fig:result_pearsonsr}
\end{figure}

\begin{figure}[!hbtp]
\centering
\includegraphics[width=1\linewidth]{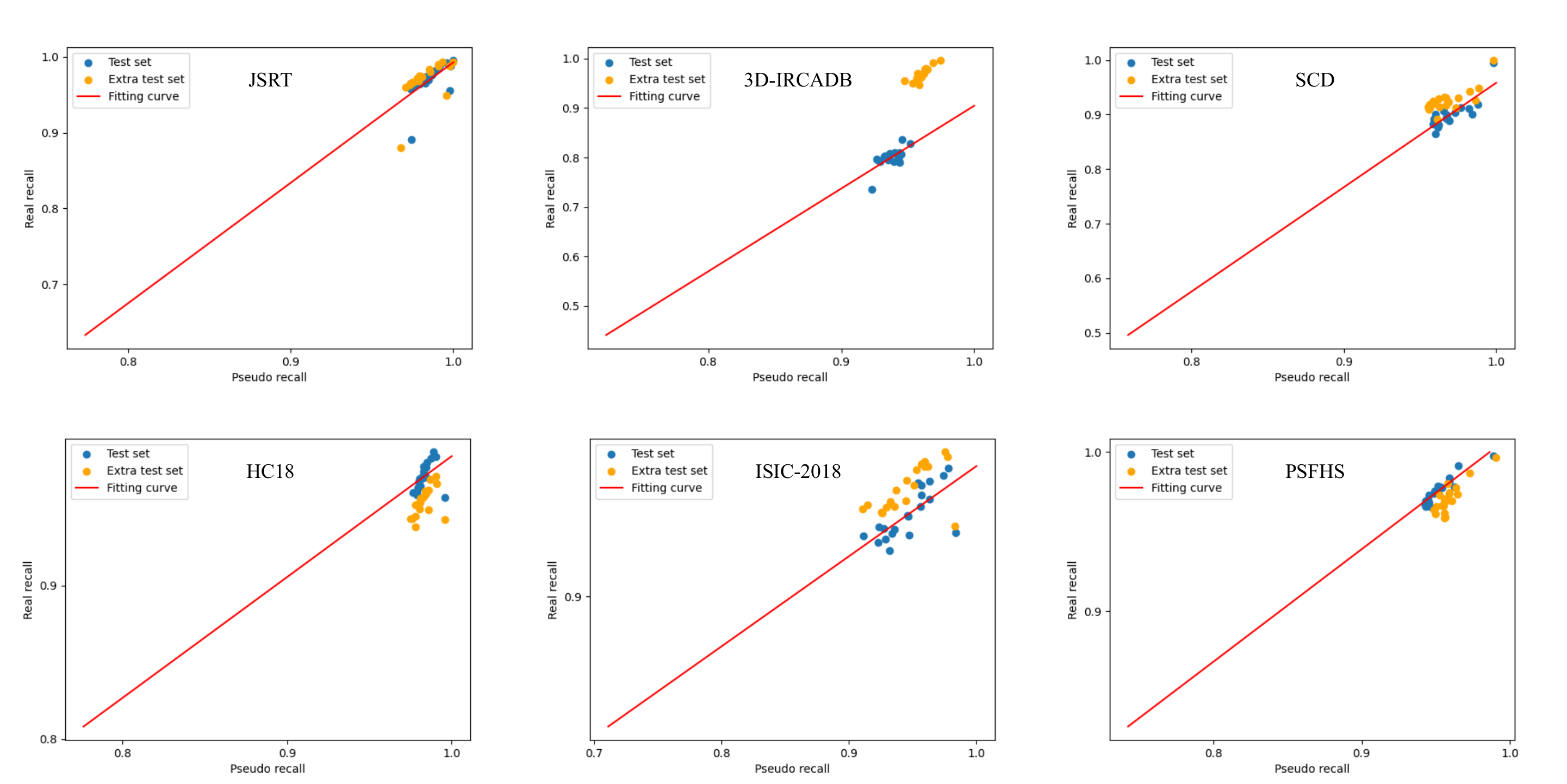}
\caption{SPE estimation of recall metrics results. }
\label{fig:result_recall}
\end{figure}

\begin{figure}[!hbtp]
\centering
\includegraphics[width=1\linewidth]{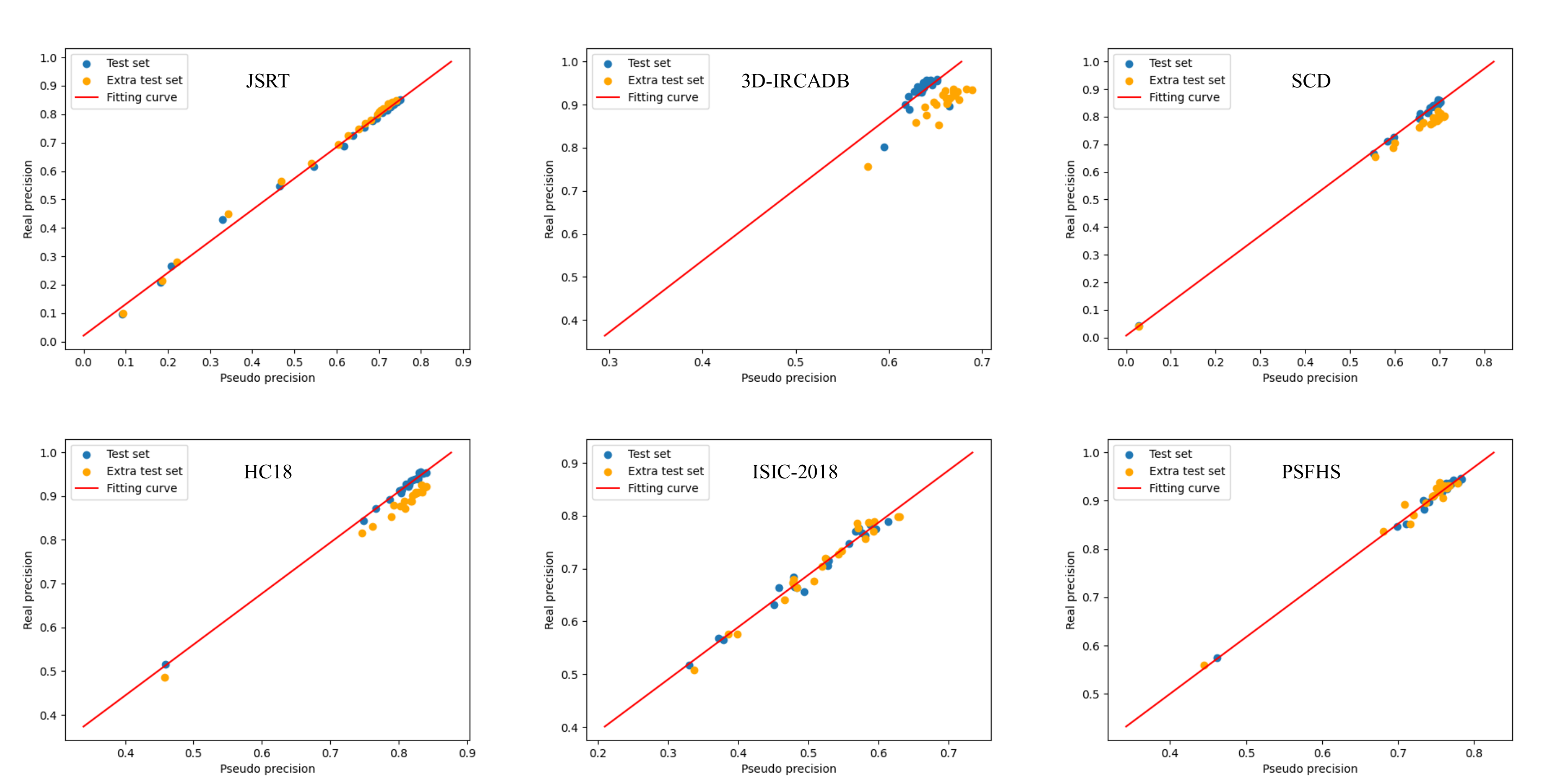}
\caption{SPE estimation of precision metrics results. }
\label{fig:result_precison}
\end{figure}

We present the result of the evaluation metrics of HD95, jaccard, Pearson correlation coefficient, recall and precision shown in Figure~\ref{fig:result_hd95},~\ref{fig:result_jaccard},~\ref{fig:result_pearsonsr},~\ref{fig:result_recall},~\ref{fig:result_precison}.

To further validate the effectiveness of SPE in estimating other metrics, we subsequently estimated the Precision metric. The experimental results are shown in Figure~\ref{fig:result_precison}. SPE performs well in estimating Precision across various datasets, especially in the JSRT, ISIC-2018 and PSFHS datasets, where the correlation reaches 0.998, 0.987, and 0.991, with MAE values of 0.014, 0.012, and 0.008, respectively, indicating that SPE can accurately estimate the Precision metric. However, the performance on the 3D-IRCADBS, SCD and HC18 datasets is slightly lower, with correlations of 0.912, 0.998 and 0.997 and MAE values of 0.063, 0.042 and 0.034. Although these errors are still small, they are slightly higher compared to other datasets. 
Overall, the performance of SPE on all datasets demonstrates its effectiveness and robustness in estimating different metrics, further supporting its potential for wide application in medical image segmentation tasks.

\newpage
\section{Understanding linear relationships of metrics for performance estimation} \label{sec:appendix:linear}
This section examines the linearity of pixel-based metrics, such as the Dice score, and distance-based metrics, like the Hausdorff distance. It further investigates why metrics like Recall and Precision exhibit similar linear characteristics to the Dice score.

Dice score, Recall, and Precision metrics focus on region overlap, emphasizing pixel-wise intersections between predicted and ground truth regions. Thus, they exhibit strong linearity. The Dice coefficient is defined as:
\begin{equation}
\text{Dice} = \frac{2 | \hat{S} \cap G |}{|\hat{S}| + |G|}
\end{equation}
where $|\hat{S}| $and $|G|$ represent the pixel counts of the predicted and ground truth regions, respectively, and $|\hat{S} \cap G|$is their intersection. Recall and Precision are defined as:
\begin{equation}
\text{Recall} = \frac{|\hat{S} \cap G|}{|G|}, \quad \text{Precision} = \frac{|\hat{S} \cap G|}{|\hat{S}|}
\end{equation}
It can be observed that all three metrics rely on the intersection $|\hat{S} \cap G|$ and the total pixel counts of the predicted region $|\hat{S}|$ and the ground truth region $|G|$.

As the predicted region gradually approaches the ground truth, the intersection $|\hat{S} \cap G|$ changes proportionally with $|\hat{S}|$ and $|G|$. For instance, when a small set of pixels $\Delta$ is added to or removed from the predicted region, the Dice coefficient can be approximated using a Taylor series expansion as:

\begin{equation}
\text{Dice} \approx \frac{2(|\hat{S} \cap G| + \Delta)}{|\hat{S}| + |G| + \Delta} \approx \frac{2|\hat{S} \cap G|}{|\hat{S}| + |G|} + \frac{2\Delta}{|\hat{S}| + |G|}
\end{equation}

Similarly, changes in Recall and Precision can be expressed as:

\begin{equation}
\Delta \text{Recall} \approx \frac{\Delta}{|G|}, \quad \Delta \text{Precision} \approx \frac{\Delta}{|\hat{S}|}
\end{equation}

These approximations demonstrate that region-overlap metrics exhibit linear responses to small changes in pixel counts, making them well-suited for SPE's linear mapping models. This linearity becomes particularly prominent when there is a high degree of overlap between the predicted and ground truth regions.

In contrast, the Hausdorff distance is defined as:

\begin{equation}
d_H(\hat{S}, G) = \max \left\{ \sup_{x \in \hat{S}} \inf_{y \in G} \|x - y\|, \sup_{y \in G} \inf_{x \in \hat{S}} \|x - y\| \right\}
\end{equation}

This metric focuses on the maximum boundary deviation between predicted and ground truth regions, making it highly sensitive to outliers. For instance, even if most boundary points have small errors, a single boundary point $x_1$that deviates significantly from its nearest ground truth boundary point $y_1$can drastically increase the Hausdorff distance. Improvements in the majority of boundary points may not proportionally reduce the distance. This extreme value amplification effect can be expressed as:

\begin{equation}
d_H = \max_i \delta_i
\end{equation}

where $\delta_i$is the distance from boundary point $x_i$to its nearest point $y_i$. Due to the asymmetric sensitivity to outliers, a linear mapping function is insufficient to accurately model the relationship between pseudo-performance and true performance for Hausdorff distance. Instead, non-linear models, such as $G(x) \approx a \log(x) + b$, may better capture this relationship.

In summary, metrics such as Dice, Recall, and Precision exhibit strong linearity due to their reliance on region overlap, making them compatible with linear mapping models for performance estimation. On the other hand, the Hausdorff distance exhibits significant non-linear characteristics due to its sensitivity to outliers, necessitating the use of non-linear mapping models to accurately capture its complex relationships.

\end{document}